\documentclass[aps,prl,twocolumn,showpacs,amsmath,amssymb,superscriptaddress,notitlepage]{revtex4-1}
\usepackage{graphicx}% Include figure files
\usepackage{dcolumn}% eqnarray table columns on decimal point
\usepackage{bm}% bold math
\newcommand{\ket}[1]{|#1\rangle}
\newcommand{\bra}[1]{\langle#1|}

\begin{document}

\title{Entanglement and spin squeezing in non-Hermitian phase transitions}
\author{Tony E. Lee}
\affiliation{ITAMP, Harvard-Smithsonian Center for Astrophysics, Cambridge, Massachusetts 02138, USA}
\affiliation{Department of Physics, Harvard University, Cambridge, Massachusetts 02138, USA}
\author{Florentin Reiter}
\affiliation{Niels Bohr Institute, University of Copenhagen, Blegdamsvej 17, DK-2100 Copenhagen, Denmark}
%\author{Monika H. Schleier-Smith}
%\affiliation{Department of Physics, Stanford University, Stanford, California 94305, USA}
\author{Nimrod Moiseyev}
\affiliation{Schulich Faculty of Chemistry and Faculty of Physics, Technion - Israel Institute of Technology, Haifa, 32000, Israel}

\date{\today}% It is always \today, today,
             %  but any date may be explicitly specified

\begin{abstract}
We show that non-Hermitian dynamics generate substantial entanglement in many-body systems. We consider the non-Hermitian Lipkin-Meshkov-Glick model and show that its phase transition occurs with maximum multiparticle entanglement: there is full N-particle entanglement at the transition, in contrast to the Hermitian case. The non-Hermitian model also exhibits more spin squeezing than the Hermitian model, showing that non-Hermitian dynamics are useful for quantum metrology. Experimental implementations with trapped ions and cavity QED are discussed.
\end{abstract}

\maketitle

Entanglement is a powerful way to understand the nature of many-body systems \cite{plenio07,horodecki09,guhne09}. Its utility has spread beyond quantum information into other areas of physics like condensed matter. In particular, it has been shown that entanglement provides new insight into condensed-matter systems and their phase transitions \cite{amico08}. Aside from fundamental interest, understanding the entanglement in condensed-matter systems allows one to use such systems for applications like quantum computing and quantum metrology \cite{pezze09,hyllus12,toth12,braunstein94}. In these applications, one would like as much entanglement as possible.

%Condensed matter and quantum information, two seemingly disparate fields, have found much cross-fertilization in recent years. Many-body physics traditionally falls under condensed matter, but it has been shown that entanglement (a staple of quantum information \cite{plenio07,horodecki09,guhne09}) provides new insight into many-body systems \cite{amico08}. Aside from fundamental interest, understanding the entanglement in condensed-matter systems allows one to use such systems for applications like quantum computing and quantum metrology \cite{pezze09,hyllus12,toth12}.

In a many-body system, a quantum phase transition changes how the particles are entangled with each other \cite{amico08}. The Lipkin-Meshkov-Glick model is the simplest model of interacting spins with a quantum phase transition, so it is an important example: the phase transition occurs with two-particle entanglement \cite{vidal04,dusuel04}, while multiparticle entanglement becomes macroscopic after the transition \cite{ma09,salvatori14}.

%Condensed matter and quantum information, two seemingly disparate fields, have found much cross-fertilization in recent years. Many-body physics traditionally falls under condensed matter, but it has been shown that entanglement (a staple of quantum information \cite{plenio07,horodecki09,guhne09}) provides new insight into many-body systems \cite{amico08}. In fact, a quantum phase transition usually coincides with a change in the way the particles are entangled with each other. For example, in the celebrated Lipkin-Meshkov-Glick model, two-particle entanglement peaks at the phase transition \cite{vidal04,dusuel04}, while multiparticle entanglement becomes macroscopic after the transition \cite{ma09}.

% Condensed matter is concerned with the behavior of complex many-body systems,

At the same time, the field of non-Hermitian quantum mechanics has drawn significant interest, especially with recent experimental results in cavities \cite{dembowski01,choi10}, waveguides \cite{ruter10}, and ultracold atoms \cite{barontini13}. The motivation is that non-Hermitian systems behave quite differently from Hermitian ones and can exhibit novel phenomena \cite{moiseyev11,berry04,heiss12,bender98,lee_other14,refael06,graefe08b,longhi09,cartarius11,wei12,wei14,zhang13,hickey13,lee14b,reiter12,chia08}. Non-Hermitian dynamics commonly arise in systems with decay or loss.
%A notable difference is that singularities already occur in \emph{finite} non-Hermitian systems \cite{moiseyev11,berry04,heiss12}. Such non-Hermitian transitions (known as exceptional points) were demonstrated in recent experiments \cite{dembowski01,choi10,ruter10}. 

\begin{figure}[b]
\centering
\begin{tabular}{cc}
\includegraphics[width=1.6 in,trim=2.7in 4.in 2.9in 4.2in,clip]{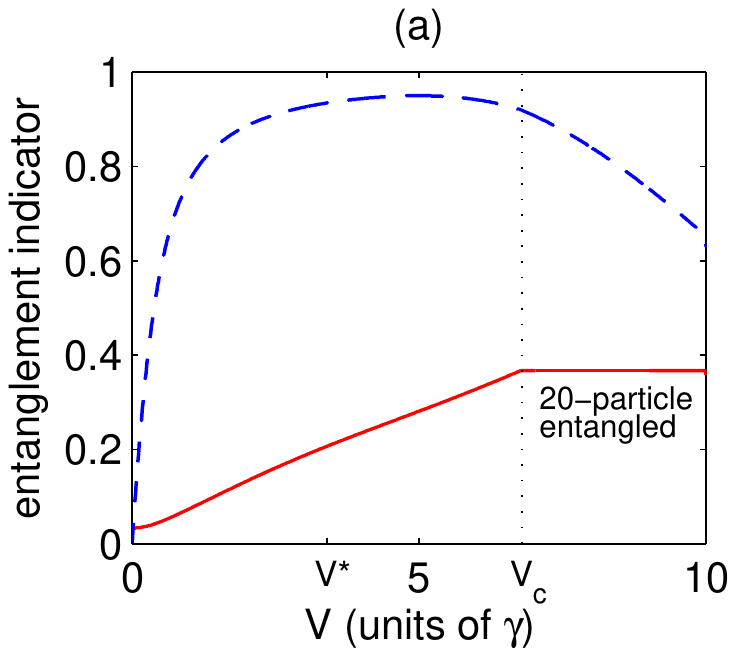}
\includegraphics[width=1.6 in,trim=3.3in 4.2in 2.6in 4.2in,clip]{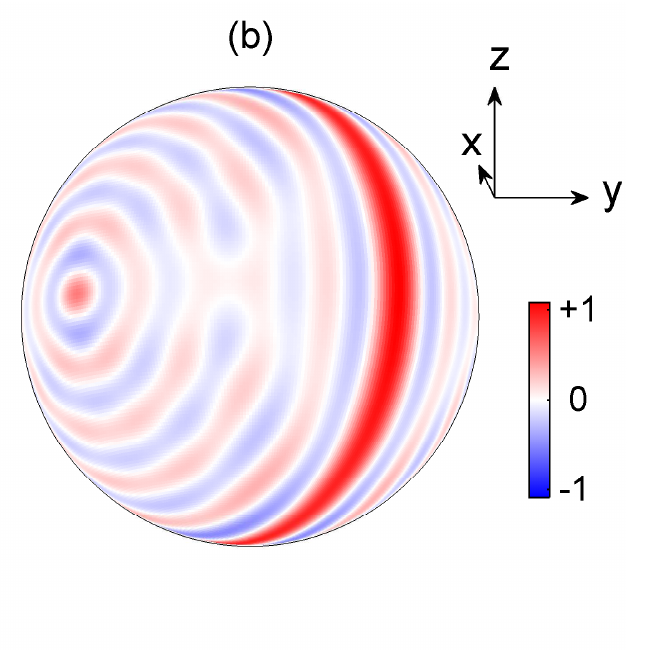}
\end{tabular}
\caption{\label{fig:first}Entanglement properties for $N=20$ spins. (a) Averaged quantum Fisher information $\bar{F}/N^2$ (solid line) indicates multiparticle entanglement, while rescaled concurrence $C_R$ (dashed line) indicates two-particle entanglement. (b) Wigner function on the Bloch sphere for $V=V_c$.}
\end{figure}

In this Letter, we view non-Hermitian quantum mechanics from a quantum-information perspective: we see what kind of entanglement it generates. We study the non-Hermitian Lipkin-Meshkov-Glick model and show that the phase transition occurs with maximum \emph{multiparticle} entanglement [Fig.~\ref{fig:first}(a)]. In fact, all particles are entangled at the transition, in contrast to the Hermitian transition. The presence of substantial multiparticle entanglement can be seen in the Wigner function, which exhibits fringes of negative value [Fig.~\ref{fig:first}(b)]. Thus, non-Hermiticity amplifies the entanglement at the phase transition.

We further show that the entanglement is useful for quantum metrology: the non-Hermitian model generates spin squeezing with phase sensitivity near the Heisenberg limit and exhibits more squeezing than the Hermitian model \cite{wineland92,kitagawa93}. Thus, non-Hermitian dynamics may be a resource for quantum metrological applications like magnetometry \cite{wasilewski10} and atomic clocks \cite{leroux10}.

%the steady state exhibits spin squeezing with phase sensitivity beyond the Hermitian model and near the Heisenberg limit

We also discuss experimental implementation with trapped ions and cavity QED. Although the scheme is probabilistic, one can implement the non-Hermitian model for thousands of atoms with a high probability, because the gap increases linearly with system size.

%, where white is zero, red is positive, and blue is negative

\emph{Model.---}
The (Hermitian) Lipkin-Meshkov-Glick model is the simplest quantum model of interacting spins \cite{botet83}. Here, we consider the non-Hermitian version,
\begin{eqnarray}
H&=&\frac{V}{N}(J_x^2 - J_y^2) - \frac{i\gamma}{2}J_z - \frac{i\gamma N}{4},\label{eq:H_spin}
\end{eqnarray}
where $\vec{J}=\frac{1}{2}\sum_n \vec{\sigma}^n$ are collective spin operators, $V$ is the coupling strength, and $N$ is the number of spins. For simplicity, we assume N is a multiple of 4. We focus on the Dicke manifold with maximum angular momentum, so the Hilbert space has dimension $N+1$. 

%Note that the length of the Bloch vector is not constant due to the non-Hermitian terms.

The Hermitian terms of Eq.~\eqref{eq:H_spin} can be experimentally implemented using trapped ions \cite{molmer99,richerme14} or cavity QED \cite{morrison08}. To obtain the non-Hermitian terms, we assume that $\left|\uparrow\right\rangle$ has a finite lifetime given by linewidth $\gamma$. Then, conditioned on the absence of a decay event, the atoms evolve with Eq.~\eqref{eq:H_spin} \cite{dalibard92,molmer93,dum92,plenio98,daley14}. In practice, one would do many experimental runs, and the runs without decay events simulate Eq.~\eqref{eq:H_spin}. The non-Hermitian evolution decreases the wavefunction norm over time due to the decrease in probability of a successful run. By having $\left|\uparrow\right\rangle$ decay into an auxiliary state instead of $\left|\downarrow\right\rangle$ and measuring the population in the auxiliary state, one can accurately determine whether a decay event occurred \cite{sherman13,myerson08}.

%; in the absence of a decay event, the atoms evolve according to Eq.~\eqref{eq:H_spin} \cite{dalibard92,molmer93,dum92,plenio98,daley14}. In practice, one would optically pump $\left|\uparrow\right\rangle$ into an auxiliary state instead of $\left|\downarrow\right\rangle$, and then measure the population in the auxiliary state to determine whether a decay event occurred \cite{sherman13,myerson08}. 

%The Hermitian terms of Eq.~\eqref{eq:H_spin} can be experimentally implemented using atoms in a cavity \cite{morrison08} or trapped ions \cite{molmer99,richerme14}. The non-Hermitian terms arise when $\left|\uparrow\right\rangle$ has a finite lifetime given by linewidth $\gamma$; in the absence of a decay event, the atoms evolve according to Eq.~\eqref{eq:H_spin} \cite{dalibard92,molmer93,dum92,plenio98,daley14}. In practice, one would optically pump $\left|\uparrow\right\rangle$ into an auxiliary state instead of $\left|\downarrow\right\rangle$, and then measure the population in the auxiliary state to determine whether a decay event occurred \cite{sherman13,myerson08}. 

%We discuss experimental considerations later in the paper.

\begin{figure}[b]
\centering
\includegraphics[width=3.5 in,trim=1.in 4.2in 1in 4.1in,clip]{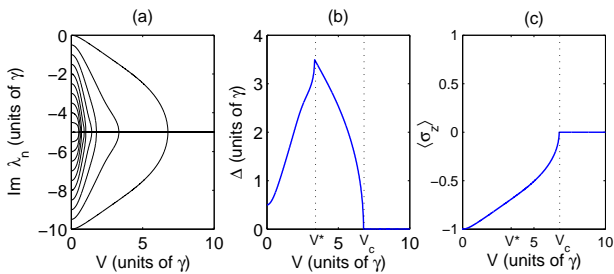}
\caption{\label{fig:eigenvalues}Eigenvalues of $H$ for $N=20$, showing (a) imaginary parts and (b) gap between the two largest imaginary parts. (c) $\langle\sigma_z\rangle$ of the steady state.}
\end{figure}

Consider the eigenvalues and eigenstates of the Hamiltonian [Eq.~\eqref{eq:H_spin}]. A wavefunction can be written as a superposition of the eigenstates of $H$. Due to the non-Hermitian terms, all eigenvalues have negative imaginary parts [Fig.~\ref{fig:eigenvalues}(a)]. Suppose one evolves a wavefunction using $\exp(-iHt)$: the weight in each eigenstate decreases over time due to the imaginary parts of the eigenvalues. After a sufficient amount of time, the wavefunction consists mostly of the eigenstate whose eigenvalue has the largest imaginary part. We are interested in this surviving eigenstate because it is the one that would be observed experimentally. We call this eigenstate the steady state since the system eventually settles into it \cite{lee12,chia08,lee14b}.

\emph{Sharp transition.---}
We are interested in whether the steady state exhibits a phase transition. We define the spectral gap $\Delta$ as the difference of the two largest imaginary parts of eigenvalues. The gap indicates how quickly the system reaches steady state. If the gap closes ($\Delta\rightarrow 0$), eigenvalues become degenerate, and the corresponding eigenstates change nonanalytically. We define $V_c$ as the value of $V$ at which the gap closes. For later usage, we define $V^*$ as the value of $V$ at which the gap is maximum.

\begin{figure}[b]
\centering
\includegraphics[width=3.3 in,trim=1.in 3.5in 1in 3.5in,clip]{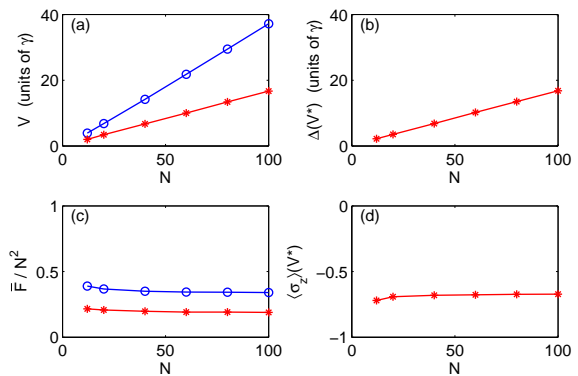}
\caption{\label{fig:scaling}Scaling of various quantities with $N$. (a) $V_c$ (blue circles) and $V^*$ (red asterisks). (b) Gap at $V^*$. (c) Averaged quantum Fisher information $\bar{F}/N^2$ at $V_c$ (blue circles) and $V^*$ (red asterisks). (d) $\langle\sigma_z\rangle$ at $V^*$.}
\end{figure}

As seen in Fig.~\ref{fig:eigenvalues}, the gap closes already for finite $N$, leading to nonanalytic behavior of $\langle\sigma_z\rangle$ at $V_c$. Non-Hermitian models are unique in their ability to have singularities for finite $N$, known as ``exceptional points'' \cite{moiseyev11,berry04,heiss12}. However, Fig.~\ref{fig:scaling}(a) shows that $V_c$ increases linearly with $N$, implying that a singularity does \emph{not} occur in an infinite system. Thus, the non-Hermitian steady state has sharp transitions for finite $N$ but not for infinite $N$; in contrast, Hermitian models have sharp transitions for infinite $N$ but not for finite $N$.

Figure \ref{fig:eigenvalues}(a) shows that there is actually a sequence of degeneracies as $V$ increases, and the degeneracy of the steady state is the last one to occur. The degeneracies can be understood by noting that the mapping $J_x,J_y,J_z\rightarrow J_y,J_x,-J_z$ leads to
\begin{eqnarray}
H + \frac{i\gamma N}{4} \rightarrow -\left(H + \frac{i\gamma N}{4}\right). \label{eq:sym}
\end{eqnarray}
This implies that the eigenvalues of $H$ are symmetric around $-i\gamma N/4$ and degenerate in pairs. 

Given the collective nature of the model, it is natural to use a mean-field approach \cite{graefe08b,lee14c}. Mean-field theory predicts that a degeneracy occurs at $V=\gamma/2$; this is actually where the first degeneracy occurs (see Supplemental Material  \cite{supplement}) and is unrelated to the steady state. Thus, the transition of the steady state (for finite $N$) is not predicted by mean-field theory.

%Figure \ref{fig:eigenvalues}(a) shows that when $V\geq V_c$, all eigenvalues have the same imaginary part, so there is no longer a unique steady state. However, all eigenstates for $V\geq V_c$ have $\langle\sigma_z\rangle=0$.

%While Hermitian systems exhibit singularities only for infinite $N$, non-Hermitian systems can exhibit singularities for finite $N$.

%Indeed, one can show this using a simple mean-field argument. 

\emph{Entanglement.---}
Having established that there is a sharp transition, we now characterize its entanglement \cite{sorensen01,guhne05}. To quantify two-particle entanglement, we use rescaled concurrence $C_R = (N-1)C$, where $C$ is the concurrence; if $C_R>0$, there is two-particle entanglement \cite{wootters98}. To quantify multiparticle entanglement, we use the averaged quantum Fisher information (QFI) \cite{hyllus12,toth12},
\begin{eqnarray}
\bar{F}&=&\frac{4}{3}[(\Delta J_x)^2 + (\Delta J_y)^2 + (\Delta J_z)^2].
\end{eqnarray}
The magnitude of $\bar{F}$ gives an indication of how much multiparticle entanglement there is; if $\bar{F}/N^2$ is on the order of 1, there is macroscopic multiparticle entanglement. In the Hermitian Lipkin-Meshkov-Glick model, rescaled concurrence peaks at the phase transition \cite{vidal04,dusuel04}, while QFI becomes macroscopic after the transition \cite{ma09,salvatori14}.

Figure \ref{fig:first}(a) shows the entanglement for the non-Hermitian model for $N=20$. (Other $N$ behave similarly.) Rescaled concurrence peaks before the transition, while QFI reaches a plateau at the transition. In fact, QFI takes the maximum possible value, $\bar{F}=(N^2+2N)/3$, when $V\geq V_c$, meaning that the steady state is fully $N$-particle entangled \cite{hyllus12,toth12}. Thus, the non-Hermitian transition is associated with \emph{multi}-particle entanglement, in contrast to the two-particle entanglement of the Hermitian transition.

To understand this behavior, we recall that if a pure symmetric state has $\langle\vec{J}\rangle=0$, it is $N$-particle entangled \cite{toth12}. At the phase transition, the steady state has $\langle\vec{J}\rangle=0$ because of three reasons. (i) $H$ is even in $J_x,J_y$, so $\langle J_x\rangle=\langle J_y\rangle=0$ always. (ii) $H+i\gamma N/4$ is $\mathcal{PT}$-symmetric \cite{graefe08a,assis09}, so its eigenvalues have 0 imaginary part at the transition (when $\mathcal{PT}$-symmetry is on the verge of breaking). (iii) $\text{Im}(H+i\gamma N/4)$ is odd in $J_z$, so $\langle J_z\rangle=0$ at the transition. Any other Hamiltonian with these three properties will also be $N$-particle entangled at its phase transition. 

% Since $H$ has spin-flip symmetry ($J_x,J_y\rightarrow -J_x,-J_y$), $\langle J_x\rangle=\langle J_y\rangle=0$. Also, due to Eq.~\eqref{eq:sym}, when the steady state degenerates, $\langle J_z\rangle=0$. Hence, at the transition, the steady state is $N$-particle entangled. (The other eigenstates are also $N$-particle entangled when they degenerate.)

% Note that the QFI is macroscopic over a broad range in $V$. Also, Fig.~\ref{fig:scaling}(c) shows that the QFI remains macroscopic as $N$ increases. 

The presence of substantial multiparticle entanglement is reflected in the Wigner function \cite{dowling94}, which exhibits interference fringes with negative values [Fig.~\ref{fig:first}(b)]. Thus, the steady state is a highly nonclassical state \cite{mcconnell13} and is similar to a rotated $|m=0\rangle$ Dicke state (see Supplemental Material \cite{supplement}). We note that Ref.~\cite{chia08} showed that the steady state of $H=-iJ_x^2$ is also a Dicke state with $N$-particle entanglement.

%The entanglement here is of Dicke type, which is why we use averaged QFI instead of nonaveraged QFI \cite{hyllus12,toth12,braunstein94}. 
% In the Supplemental Material, we show that the entanglement is of Dicke type by comparing the steady state with a rotated Dicke state.

%so the Wigner function does not resemble a ``Schr\"{o}dinger's cat'' \cite{mcconnell13}

%Figure \ref{fig:first}(a) shows that the QFI is large even when $V<V_c$. For example, when $V=V^*$, there is still 13-particle entanglement. The presence of substantial multiparticle entanglement is reflected in the Wigner function \cite{dowling94}, which exhibits interference fringes with negative values [Fig.~\ref{fig:first}(b)]. Thus, the steady state is a highly nonclassical state.

Figure \ref{fig:first}(a) shows that even when $V<V_c$, QFI remains large, meaning that there is still a lot of multiparticle entanglement. For example, when $V=V^*$, there is still 13-particle entanglement \cite{hyllus12,toth12}.

%Note that the Wigner function here is not a ``Schr\"{o}dinger cat'' state.

%For example, for $N=20$, there is 16-particle entanglement at $V=V_c$. 
%We also define $V^*$ as the value of $V$, at which the gap is a maximum.

\emph{Spin squeezing.---}
Now we show that the entanglement is useful for quantum metrology by calculating the spin squeezing of the steady state. When an ensemble of atoms is spin squeezed, one can measure rotations on the Bloch sphere better than the shot-noise limit, which is important for precision measurements. We use the spin-squeezing parameter as defined by Wineland \emph{et al.}~\cite{wineland92},
\begin{eqnarray}
\xi^2&=&\min_{\vec{n}_\perp} \, \frac{N(\Delta J_{\vec{n}_\perp})^2}{|\langle \vec{J} \rangle|^2}, \label{eq:xi2}
\end{eqnarray}
where $\vec{n}_\perp$ is a unit vector normal to $\langle \vec{J} \rangle$. There is squeezing when $\xi^2<1$; the smaller $\xi^2$ is, the better the phase sensitivity.

\begin{figure}[t]
\centering
\includegraphics[width=3.2 in,trim=1.2in 4.2in 1.2in 4.2in,clip]{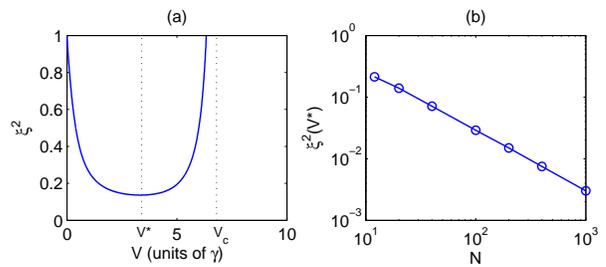}
\caption{\label{fig:squeezing}(a) Spin squeezing for $N=20$ as function of $V$. (b) Minimum $\xi^2$ for different $N$.}
\end{figure}

Figure \ref{fig:squeezing}(a) shows that $\xi^2$ reaches a minimum at $V^*$, which is where the gap is maximum [Fig.~\ref{fig:eigenvalues}(b)]. Figure \ref{fig:squeezing}(b) shows the squeezing for different $N$ and indicates $\xi^2\approx 3/N$, so the phase sensitivity is near the Heisenberg limit ($\xi^2= 1/N$). 

%Squeezing is evident in the Wigner function [Fig.~\ref{fig:first}(b)].

For comparison, squeezing of the Hermitian ground state scales as $\xi^2\sim N^{-1/3}$ \cite{dusuel04}. Time evolution with the Hermitian Hamiltonian (two-axis countertwisting model \cite{kitagawa93}) leads to squeezing with $\xi^2\approx 4/N$. Thus, the non-Hermitian model has \emph{more} squeezing than the Hermitian model. It also surpasses the master equation's steady state ($\xi^2=1/2$) \cite{lee13c}.

We note that there are other measurement-based spin-squeezing protocols, starting from Kuzmich \emph{et al.} \cite{kuzmich98,saffman09,norris12}. Our scheme uses a different type of measurement (absence of a decay event), which leads to the explicit non-Hermitian Hamiltonian in Eq.~\eqref{eq:H_spin}. This non-Hermitian scheme may be advantageous in situations where the decay of $\left|\uparrow\right\rangle$ is non-negligible. Also, since the scheme is based on a steady state, it is robust to initial conditions.

%Squeezing is also evident in the Wigner function [Fig.~\ref{fig:first}(b)]. As $V$ approaches $V_c$, the squeezing disappears because the Bloch vector shrinks in length [Fig.~\ref{fig:eigenvalues}(c)]. (Note that $\langle\sigma_x\rangle=\langle\sigma_y\rangle=0$.)

%Figure \ref{fig:squeezing}(b) shows the squeezing for different $N$ and indicates the scaling, $\xi^2=3/N$, meaning that the phase sensitivity is near the Heisenberg limit ($\xi^2= 1/N$). Note that for the Hermitian ground state, the squeezing is not Heisenberg-limited: $\xi^2\sim N^{-1/3}$ \cite{dusuel05}.

% This is a fortunate coincidence: since the gap is maximum at $V=V^*$, the system reaches steady state the fastest there.

\emph{Probabilities.---}
The non-Hermitian scheme is probabilistic, since it is conditioned on the absence of a decay event among $N$ atoms. An important question is how scalable the scheme is: for large $N$, what is the probability that an experimental trial reaches steady state before a decay event? One expects that as $N$ increases, the probability should decrease exponentially. This turns out to be wrong due to two fortunate coincidences.

The time to reach steady state is on the order of $1/\Delta$. The average number of decay events during this time is \cite{dalibard92}
\begin{eqnarray}
\mu&=& \frac{\gamma N(\langle\sigma_z\rangle + 1)}{2\Delta}.
\end{eqnarray}
The probability of no decay event is $e^{-\mu}$.

%The probability of a successful trial is the probability of no decay event, which is $e^{-\mu}$.

%The rate of decay events for each atom is $\gamma(\langle\sigma_z\rangle+1)/2$.

It is advantageous to set $V=V^*$, since $\Delta$ is maximum and $\xi^2$ is minimum there. Now, it turns out that $\Delta(V^*)$ increases linearly with $N$ [Fig.~\ref{fig:scaling}(b)]. To estimate $\langle\sigma_z\rangle$, we use its steady-state value, which is independent of $N$ when $V=V^*$ [Fig.~\ref{fig:scaling}(d)]. Thus, this rough estimate says that the probability of success is \emph{independent} of $N$.

\begin{figure}[t]
\centering
\includegraphics[width=3.2 in,trim=1.1in 4.3in 1.2in 4.3in,clip]{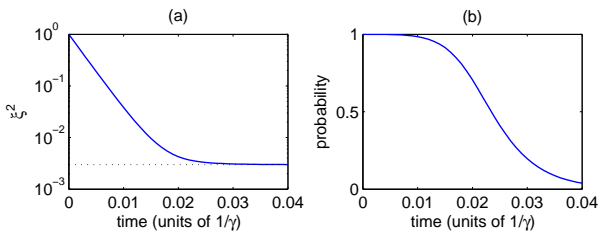}
\caption{\label{fig:probability}Non-Hermitian evolution of $N=1000$ spins with $V=V^*$, starting from $\left|\downarrow\downarrow\ldots\downarrow\right\rangle$. (a) Spin squeezing at current time (solid line) and steady state (dotted line). (b) Probability of no decay event.}
\end{figure}

For a more accurate estimate, Fig.~\ref{fig:probability} shows the non-Hermitian evolution of $N=1000$ spins starting with all spins in $\left|\downarrow\right\rangle$. As time increases, $\xi^2$ decreases towards the steady-state value, and the probability of no decay event decreases. The squeezing reaches steady state at a time of about $0.025/\gamma$, which corresponds to a probability of $0.4$. This clearly shows that the non-Hermitian scheme is feasible for a large number of spins. This is due to two fortunate coincidences: $\xi^2$ is minimum when $\Delta$ is maximum, and $\Delta$ increases linearly with $N$.

\emph{Bosonic approximation.---}
The above results were obtained numerically using exact diagonalization. One can obtain many results analytically using the Holstein-Primakoff transformation. We expand around $J_z=-N/2$ by mapping $J_z\rightarrow-N/2 + a^\dagger a$ and $J_-\rightarrow\sqrt{N}a$, where $a^\dagger$ and $a$ are bosonic creation and annihilation operators that satisfy $[a,a^\dagger]=1$. This mapping is accurate when $a^\dagger a\ll N$. Equation \eqref{eq:H_spin} becomes 
\begin{eqnarray}
H&=&\frac{V}{2}(a^{\dagger2}+a^2) - \frac{i\gamma}{2}a^\dagger a \label{eq:H_boson},
\end{eqnarray}
which can be diagonalized using a complex Bogoliubov transformation \cite{zhang13,hickey13,lee14b}:
\begin{eqnarray}
H&=&-\frac{i}{2}\sqrt{4V^2+\gamma^2}\; \bar{b}b - \frac{i}{4}(\sqrt{4V^2+\gamma^2}-\gamma), \quad \label{eq:H_diag}\\
%a^\dagger&=&\bar{b}\,\cosh\frac{\theta}{2}   + b \, \sinh\frac{\theta}{2},\\
%a&=&\bar{b}\,\sinh\frac{\theta}{2}  + b \, \cosh\frac{\theta}{2},
a^\dagger&=&\bar{b}\,\cosh\frac{\theta}{2}   + b \, \sinh\frac{\theta}{2}, \quad\quad  a=\bar{b}\,\sinh\frac{\theta}{2}  + b \, \cosh\frac{\theta}{2},\nonumber
%-\frac{2iV}{\gamma}&=&\tanh\theta, \label{eq:theta}
\end{eqnarray}
where $\theta$ satisfies $-2iV/\gamma=\tanh\theta$, and $\bar{b}$ and $b$ are bosonic creation and annihilation operators that satisfy $[b,\bar{b}]=1$. It is important to realize that $\bar{b}\neq b^\dagger$ because $\theta$ is complex. The vacuum state of the $b$ bosons is defined via $b|0\rangle=0$. We identify $|0\rangle$ as the steady state because its eigenvalue has the largest imaginary part.

The eigenvalues are given by Eq.~\eqref{eq:H_diag}, and the bosonic model never has a degeneracy. We recall that the original model has eigenvalues symmetric around $-i\gamma N/4$ [Eq.~\eqref{eq:sym}]. Equation \eqref{eq:H_diag} predicts only the eigenvalues above $-i\gamma N/4$. To get the other eigenvalues, we have to expand around $J_z=N/2$. The symmetry implies that a degeneracy occurs when an eigenvalue reaches $-i\gamma N/4$.

%their imaginary parts decrease as $V$ increases

This allows us to predict, for large $N$ (see Supplemental Material \cite{supplement}),
\begin{eqnarray}
V^*&=&\Delta(V^*)=\frac{\gamma N}{6}, \quad \langle\sigma_z\rangle(V^*) = -\frac{2}{3}, \label{eq:pred1}\\
  V_c&=&\frac{\gamma N}{2}, \quad\xi^2(V^*)=\frac{27}{8N}, \quad\frac{\bar{F}(V^*)}{N^2}=\frac{8}{27}. \label{eq:pred2}
\end{eqnarray}
Equations \eqref{eq:pred1} are surprisingly accurate, while Eqs.~\eqref{eq:pred2} have the right scaling with $N$ but not the right prefactor.

\emph{Experimental considerations.---}
The Hermitian part of Eq.~\eqref{eq:H_spin} can be implemented using trapped ions \cite{molmer99} or atoms in a cavity \cite{morrison08}. A recent experiment implemented a similar model with 11 ions and $V\sim \text{1 kHz}$ \cite{richerme14}. To get the non-Hermitian terms, one would optically pump from $\left|\uparrow\right\rangle$ into an auxiliary state so that $\left|\uparrow\right\rangle$ has linewidth $\gamma$ (see Supplemental Material \cite{supplement}). By measuring the population in the auxiliary state, one can determine with near perfect efficiency whether a decay event occurred \cite{sherman13,myerson08}. One would do multiple experimental runs, and the runs without decay events are the ones that simulate the non-Hermitian model. The non-Hermitian evolution was experimentally demonstrated with one ion \cite{sherman13}. Thus, the experimental implementation of Eq.~\eqref{eq:H_spin} is well within current technology. 

%The Hermitian terms of Eq.~\eqref{eq:H_spin} can be implemented using atoms in cavities \cite{morrison08} or trapped ions \cite{molmer99,richerme14}: the spin-spin interaction is mediated by the cavity or collective ion motion. To get the non-Hermitian terms, one would optically pump from $\left|\uparrow\right\rangle$ into an auxiliary state so that $\left|\uparrow\right\rangle$ has linewidth $\gamma$. By measuring the population in the auxiliary state, one can determine with near perfect efficiency whether a decay event occurred \cite{sherman13,myerson08}. One would do multiple experimental runs, and the runs without decay events are the ones that simulate the non-Hermitian model. The Supplemental Material provides suitable optical-pumping schemes for ${}^{87}\text{Rb}$ and ${}^{43}\text{Ca}^+$.

To see the sharp transition, one would look for the singularity of $\langle\sigma_z\rangle$ as a function of $V$ [Fig.~\ref{fig:eigenvalues}(c)]. When $V<V_c$, there is a unique steady state, and each experimental run should last for a time of at least $1/\Delta$ to reach steady state. When $V>V_c$, there is not a unique steady state, but all eigenstates have $\langle\sigma_z\rangle=0$, which can be observed by averaging over time. Note that the relevant parameter is $V/\gamma$, which can be made large by setting $\gamma$ small.

%The entanglement, spin squeezing, and Wigner function can be found via correlation measurements of the atoms. 
%Since $\Delta\rightarrow 0$ as $V\rightarrow V_c$, one would not work very close to $V_c$. 

\emph{Conclusion.---}
We have shown that quantum information sheds new light on non-Hermitian many-body systems. Non-Hermitian dynamics can amplify the entanglement and spin squeezing near quantum phase transitions. One should consider other non-Hermitian models to see how general this is. In particular, it would be interesting to see the effect of non-Hermiticity on topological entanglement entropy \cite{kitaev06,levin06}. Finally, one should study how non-Hermitian terms affect the entanglement scaling in one-dimensional spin chains \cite{osterloh02,osborne02,vidal03,hu13,joshi13}.

We thank Monika Schleier-Smith, Ching-Kit Chan, Raam Uzdin, and Swati Singh for useful discussions. This work was supported by the NSF through a grant to ITAMP. N.M. acknowledges ICore: the Israeli Excellence Center ``Circle of Light'' for partial support. F.R. acknowledges support from the Studienstiftung des deutschen Volkes.

\bibliography{nonherm_lmg}

\newpage
\appendix
\setcounter{figure}{5}    
\setcounter{equation}{12}

\onecolumngrid

\section{SUPPLEMENTAL MATERIAL}

%In this supplement, we provide more details about experimental implementation and non-Hermitian degeneracies.

\section{Non-Hermitian degeneracies (exceptional points)}

\begin{figure}[h]
\centering
\includegraphics[width=5.5 in,trim=0in 3.3in 0.3in 3.5in,clip]{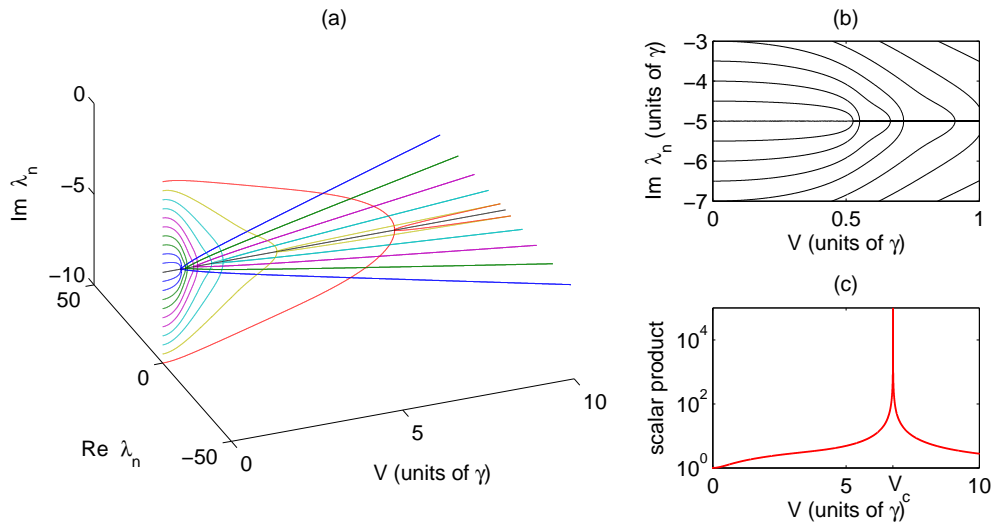}
\caption{\label{fig:eigenvalues_3d}Non-Hermitian features for $N=20$. (a) Real and imaginary parts of eigenvalues of $H$. (b) Zoomed-in view of imaginary parts of eigenvalues. (c) Scalar product of the steady state.}
\end{figure}

Figure \ref{fig:eigenvalues_3d}(a) plots both the real and imaginary parts of the eigenvalues of $H$ as a function of $V$ for $N=20$. There is a sequence of $N/2$ degeneracies as $V$ increases. Before a degeneracy of an eigenvalue pair, both eigenvalues are purely imaginary; after the degeneracy, they are complex with imaginary part $-i\gamma N/4$. Another way of saying this is that $H+i\gamma N/4$ is $\mathcal{PT}$-symmetric \cite{graefe08a}, and $\mathcal{PT}$-symmetry is broken when $V<V_c$.

Figure \ref{fig:eigenvalues_3d}(b) is a zoomed-in view of the imaginary parts of the eigenvalues. The first degeneracy occurs close to $V=\gamma/2$, which is the critical point predicted by mean-field theory. As $N$ increases, the degeneracy moves closer to $V=\gamma/2$.

Figure \ref{fig:eigenvalues_3d}(c) demonstrates the phenomenon of self-orthogonality. Let the right eigenvectors of $H$ be denoted $u_n$. Since $H$ is non-Hermitian, the eigenvectors are normalized and orthogonal according to the c-product, $u_n^T \cdot u_m = \delta_{mn}$, which is different from the usual scalar product, $u_n^{\dagger} \cdot u_m = \delta_{mn}$. It turns out that at an exceptional point, the eigenvector becomes self-orthogonal, $u_n^T \cdot u_n = 0$. Numerically, this is seen as a divergence in the scalar product. Indeed, Fig.~\ref{fig:eigenvalues_3d}(c) shows that the scalar product of the steady state diverges at the exceptional point. For more information, see Chapter 9 of Ref.~\cite{moiseyev11}.

Interestingly, at a non-Hermitian degeneracy, the two eigenvectors become parallel. Also, the survival probability develops a linear dependence on time \cite{cartarius11}.

\section{Comparison with rotated Dicke state}

Here, we show that the multiparticle entanglement of the non-Hermitian steady state is of Dicke-type by showing that the steady state is similar to a rotated Dicke state. Let $|\psi\rangle$ be the steady state as a function of $V$. (Strictly speaking, when $V\geq V_c$, there is not a unique steady state but we continuously follow the eigenstate that is the unique steady state for $V<V_c$.) Let $|\psi'\rangle$ be the Dicke state $|j=N/2,m=0\rangle$ rotated by angle $\pi/2$ around the axis $(\hat{x}+\hat{y})/\sqrt{2}$:
\begin{eqnarray}
|\psi'\rangle&=&\exp\left(\frac{-i\pi(J_x+J_y)}{2\sqrt{2}}\right) |j=N/2,m=0\rangle.
\end{eqnarray}

\begin{figure}[h]
\centering
\begin{tabular}{cc}
\includegraphics[width=4in,trim=1.in 4.2in 1.2in 4.2in,clip]{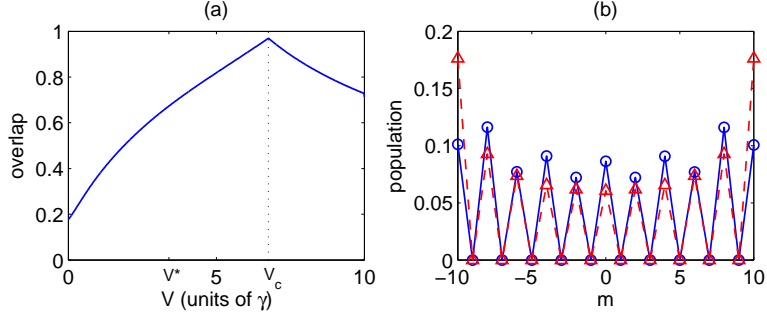}
\end{tabular}
\caption{\label{fig:overlap}Comparison of steady state with rotated Dicke state for $N=20$. (a) Overlap $|\langle\psi'|\psi\rangle|^2$ as a function of $V$. (b) Population in each $m$ component: $|\langle m|\psi\rangle|^2$ at $V_c$ (blue circles, solid line) and $|\langle m|\psi'\rangle|^2$ (red triangles, dashed line).}
\end{figure}

Figure \ref{fig:overlap}(a) shows the overlap  $|\langle\psi'|\psi\rangle|^2$ as a function of $V$ for $N=20$. The overlap is maximum (0.97) at $V_c$. Figure \ref{fig:overlap}(b) shows the population in each $m$ component for $|\psi\rangle$ at $V_c$ and $|\psi'\rangle$. They are clearly similar. Thus, at the transition, the steady state is similar to (but not exactly) a rotated Dicke state.

Note that $|\psi'\rangle$ has maximum averaged quantum Fisher information, $\bar{F}=(N^2+2N)/3$, so it is $N$-particle entangled \cite{hyllus12,toth12}.

\section{Bosonic model: expectation values}

Here, we provide details on calculating expectation values in the bosonic model. It is more convenient to express $a,a^\dagger$ in terms of $b,b^\dagger$ instead of $b,\bar{b}$:
\begin{eqnarray}
a^\dagger = \frac{ b^\dagger \cosh\frac{\theta}{2} +  b \left(\sinh\frac{\theta}{2}\right)^*}{\left|\cosh\frac{\theta}{2}\right|^2 - \left|\sinh\frac{\theta}{2}\right|^2} ,
\quad\quad\quad
a = \frac{ b^\dagger\sinh\frac{\theta}{2} + b \left(\cosh\frac{\theta}{2}\right)^* }{\left|\cosh\frac{\theta}{2}\right|^2 - \left|\sinh\frac{\theta}{2}\right|^2}. 
\label{aofbbdagger}
\end{eqnarray}
We also express $b^\dagger$ in terms of $b,\bar{b}$:
\begin{eqnarray}
b^\dagger = \left(\left|\cosh\frac{\theta}{2}\right|^2 - \left|\sinh\frac{\theta}{2}\right|^2\right) \bar{b} + \left( \left( \cosh\frac{\theta}{2} \right)^* \sinh\frac{\theta}{2} - \cosh\frac{\theta}{2} \left( \sinh\frac{\theta}{2} \right)^* \right) b,
\label{bdaggerofbbbar}
\end{eqnarray}
whereby we find
\begin{eqnarray}
\bra{0} b b^\dagger \ket{0} = \left|\cosh\frac{\theta}{2}\right|^2 - \left|\sinh\frac{\theta}{2}\right|^2.
\end{eqnarray}

We take expectation values with respect to the vacuum of the $b$-bosons, since it is the steady state. First,
\begin{eqnarray}
\langle a^{\dagger 2} \rangle
&=& \frac{\cosh\frac{\theta}{2} \left(\sinh\frac{\theta}{2}\right)^*}{\left(\left|\cosh\frac{\theta}{2}\right|^2 - \left|\sinh\frac{\theta}{2}\right|^2\right)^2} \bra{0} b b^\dagger \ket{0}
= \frac{\cosh\frac{\theta}{2} \left(\sinh\frac{\theta}{2}\right)^*}{\left|\cosh\frac{\theta}{2}\right|^2 - \left|\sinh\frac{\theta}{2}\right|^2},
\\
\langle a^2 \rangle
&=& \frac{\sinh\frac{\theta}{2} \left(\cosh\frac{\theta}{2}\right)^*}{\left(\left|\cosh\frac{\theta}{2}\right|^2 - \left|\sinh\frac{\theta}{2}\right|^2\right)^2} \bra{0} b b^\dagger \ket{0}
= \frac{\sinh\frac{\theta}{2} \left(\cosh\frac{\theta}{2}\right)^*}{\left|\cosh\frac{\theta}{2}\right|^2 - \left|\sinh\frac{\theta}{2}\right|^2},
\\
\langle a^\dagger a \rangle
&=& \frac{\left|\sinh\frac{\theta}{2}\right|^2}{\left(\left|\cosh\frac{\theta}{2}\right|^2 - \left|\sinh\frac{\theta}{2}\right|^2\right)^2} \bra{0} b b^\dagger \ket{0}= \frac{\left|\sinh\frac{\theta}{2}\right|^2}{\left|\cosh\frac{\theta}{2}\right|^2 - \left|\sinh\frac{\theta}{2}\right|^2}.
\end{eqnarray}
We use hyperbolic identities to obtain 
\begin{eqnarray}
\sinh\frac{\theta}{2} &=& \sqrt{\frac{1}{2} \left(\cosh\theta - 1\right)} = \sqrt{\frac{1}{2} \left(\frac{\gamma}{\sqrt{4 V^2 + \gamma^2}} - 1\right)} = -i \sqrt{\frac{1}{2} \left(1 - \frac{\gamma}{\sqrt{4 V^2 + \gamma^2}}\right)}, \label{eq:sinh}
\\
\cosh\frac{\theta}{2} &=& \sqrt{\frac{1}{2} \left(\cosh\theta + 1\right)} = \sqrt{\frac{1}{2} \left(\frac{\gamma}{\sqrt{4 V^2 + \gamma^2}} + 1\right)}.
\end{eqnarray}
Since Im $\tanh\theta < 0$, we have to let $\sinh\frac{\theta}{2}$ be negative in the last step of Eq.~\eqref{eq:sinh}. This gives
\begin{eqnarray}
\bra{0} b b^\dagger \ket{0} &=& \left|\cosh\frac{\theta}{2}\right|^2 - \left|\sinh\frac{\theta}{2}\right|^2 = \cosh\theta = \frac{\gamma}{\sqrt{4 V^2 + \gamma^2}},
\\
\langle  a^{\dagger 2} \rangle
&=& +\frac{i}{2} | \tanh\theta | = + \frac{i V}{\gamma},
\label{adaggersquaredexpres}
\\
\langle a^2 \rangle
&=& -\frac{i}{2} | \tanh\theta | = -\frac{i V}{\gamma},
\label{asquaredexpres}
\\
\langle a^\dagger a \rangle
&=& \frac{1 - \cosh\theta}{2 \cosh\theta} = \frac{1}{2} \left( \frac{\sqrt{4 V^2 + \gamma^2}}{\gamma} - 1 \right).
\label{adaggeraexpres}
\end{eqnarray}

From these results, we find:
\begin{eqnarray}
\langle\sigma_z\rangle &=& -1 + \frac{-1+\sqrt{4(V/\gamma)^2+1}}{N}, \\
\xi^2&=& \frac{N^2\left[-2V+\sqrt{4(V/\gamma)^2+1}\right]}{\left[N+1-\sqrt{4(V/\gamma)^2+1}\right]^2},\\
\bar{F}&=&\frac{2}{3}\left[N\sqrt{4(V/\gamma)^2+1} + 4(V/\gamma)^2\right].
\end{eqnarray}

\section{Experimental level schemes}

\begin{figure}[h]
\centering
\includegraphics[width=5 in,trim=0in 6in 0in 0in,clip]{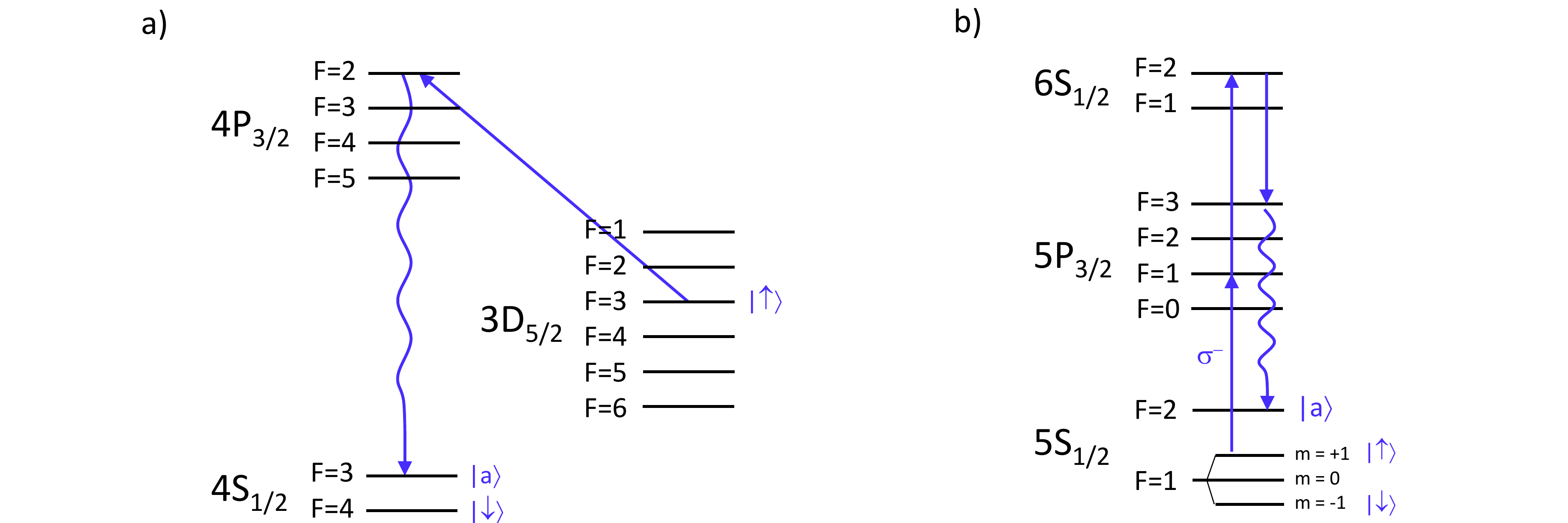}
\caption{\label{fig:experiment}Optical-pumping schemes for (a) ${}^{43}\text{Ca}^+$ and (b) ${}^{87}\text{Rb}$.}
\end{figure}

As discussed in Ref.~\cite{sherman13}, the optical pumping should be such that $\left|\uparrow\right\rangle$ decays mostly into an auxiliary state $\left|a\right\rangle$ instead of $\left|\downarrow\right\rangle$. It is advantageous to use atoms with hyperfine structure since they have many ground states. Figure \ref{fig:experiment} shows suitable level schemes for ${}^{43}\text{Ca}^+$ and ${}^{87}\text{Rb}$.

\end{document}